# A New Formalism for Numerical Relativity

Carles Bona[1], Joan Massó[1,2], Edward Seidel[2,3] and Joan Stela[1]

[1] *Departament de Física, Universitat de les Illes Balears, E-07071 Palma de Mallorca, SPAIN*
[2] *National Center for Supercomputer Applications, 605 East Springfield Avenue, Champaign, IL 61280*
[3] *Department of Physics, University of Illinois, Urbana, IL 61801*

We present a new formulation of the Einstein equations that casts them in an explicitly first order, flux-conservative, hyperbolic form. We show that this now can be done for a wide class of time slicing conditions, including maximal slicing, making it potentially very useful for numerical relativity. This development permits the application to the Einstein equations of advanced numerical methods developed to solve the fluid dynamic equations, *without* overly restricting the time slicing, for the first time. The full set of characteristic fields and speeds is explicitly given.

PACS numbers: 04.25.Dm

*Introduction.* For decades the standard approach [1] to numerical relativity has been based on a direct application of the 3+1 formulation of the Einstein equations by Arnowitt, Deser, and Misner [2]. This important contribution laid the foundation for most numerical work in the field. As convenient as this formulation is for numerical relativity, the structure of the equations is extremely complicated and most of the work consisted in developing "ad hoc" numerical codes for every case considered. This is in contrast with the situation of modern Computational Fluid Dynamics, which deals with first order, flux conservative, hyperbolic (FOFCH) systems of equations, for which many advanced numerical methods have been developed [3] based on the particular mathematical structure of the equations [4].

After many pioneering attempts [5–7] it was shown that the full set of 3D Einstein equations could be put in the FOFCH form [8], similar to hydrodynamics. This development allowed the same advanced numerical methods used in hydrodynamics, such as conservative schemes and modern shock capturing methods, to be applied to the Einstein equations for the first time. In a number of tests involving spherical black holes [9,10], 1D general relativistic hydrodynamics [11] and 3D gravitational waves [12], this formulation showed some of its strengths over the standard approach. However, the price to pay for this original formulation was that it required the restrictive assumption of harmonic time slicing. Although this slicing condition is singularity avoiding, making it potentially useful for studies of strongly gravitating systems, it is just barely so [13]. For this reason metric and curvature components can grow without bound near the singularity and, without some sort of horizon boundary condition [10,14,15], this slicing is not well suited for black hole spacetimes.

In this Letter, we will show that this FOFCH formalism can now be extended to a rather wide class of slicing conditions suitable for many different strong and weak field applications, *including* the time honored choice of maximal slicing. This new development opens the door to the use of very mature numerical methods in numerical relativity for many problems of interest, including black hole spacetimes. It should also facilitate mathematical studies of the Einstein Equations, as there is a vast literature on systems of hyperbolic conservation laws of this form (see, e.g., Refs. [3,4]) that can now be applied to general relativity.

*Evolution equations.* It is well known that Einstein's field equations can be decomposed into two sets: the evolution system and the constraints. The energy and momentum constraints contain no second time derivatives. The remaining equations form the evolution system. This is just a matter of choice, because an evolution equation plus a constraint leads to another evolution equation with the same physical solutions (the ones obtained from initial data which satisfy the constraints).

The standard choice for the evolution system is to take the space components of the Ricci tensor, but one could choose instead the space components of the Einstein tensor or any other combination obtained by using the constraints in a suitable way. The freedom to choose the coordinate gauge allows one to complete the evolution system in many different ways, and this can also lead to many different systems of equations, each one with its own structure. Nevertheless, we know that the physical solutions of all these systems (the ones obtained from initial data which satisfy the constraints) are equivalent, although some systems will be better suited for numerical studies. Below, we will use special choices of variables, and make use of the constraints in the evolution system and the gauge choice to derive a system with the FOFCH structure.

The standard 3+1 ADM evolution system is given by:

$$(\partial_t - \mathcal{L}_\beta)\gamma_{ij} = -2\alpha\, K_{ij} \tag{1}$$

and the evolution equations:

$$(\partial_t - \mathcal{L}_\beta)K_{ij} = -\alpha_{;j} + \alpha\,[R^{(3)}_{ij} + tr\, K\, K_{ij} - 2K^{ij}K_{ij} - R^{(4)}_{ij}]\,, \tag{2}$$

where indices are raised with the inverse matrix $\gamma^{ij}$ of the space metric. This system is first order in time, but second order in space. To obtain a system which is also first order in space, we will introduce auxiliary variables which correspond to the space derivatives,

$$A_k = \partial_k ln\,\alpha\,,\quad B_k^{\ i} = 1/2\,\partial_k \beta^i\,,\quad D_{kij} = 1/2\,\partial_k \gamma_{ij}\,. \tag{3}$$

Note that $B$, $D$ or the shift $\beta$ are not tensor quantities on the constant time hypersurfaces. Nevertheless, we will formally raise and lower indices with the three-metric $\gamma_{ij}$.

One could then simply insert these quantities into the standard ADM equations to obtain a first order system. However, doing so blindly does not bring any particular advantage. A careful choice of variables will transform the equations into the FOFCH form that is especially suited to mathematical analysis and numerical treatment. In particular, the evolution system can be written as

$$\partial_t K_{ij} + \partial_r[-\beta^r K_{ij} + \alpha\,\lambda^r_{ij}] = \alpha\, S_{ij}\,. \tag{4}$$

where the terms $\lambda^k_{ij}$ are given by

$$\lambda^k_{ij} = D^k_{\ ij} + 1/2\,\delta^k_i\,(A_j + 2V_j - D_{j\ r}^{\ r}) + 1/2\,\delta^k_j\,(A_i + 2V_i - D_{i\ r}^{\ r})\,, \tag{5}$$

and we have noted

$$V_k = D_{kr}^{\ \ r} - D^r_{\ rk}\,. \tag{6}$$

$S_{ij}$ is a source term involving *only* the fields themselves and *not* their derivatives:

$$\begin{aligned}S_{ij} = &-R^{(4)}_{ij} - 2K_i^{\ k}K_{kj} + tr\, K\, K_{ij}\\&+ 2/\alpha\,(K_{ir}B_j^{\ r} + K_{jr}B_i^{\ r})\\&+ 4D_{kri}D^{kr}_{\ \ j} + \Gamma^k_{\ kr}\Gamma^r_{\ ij} - \Gamma_{ikr}\Gamma_j^{\ kr}\\&- (2D^{kr}_{\ \ k} - A^r)(D_{ijr} + D_{jir})\\&+ A_i(V_j - 1/2\,D_{j\ k}^{\ k}) + A_j(V_i - 1/2\,D_{i\ k}^{\ k})\,.\end{aligned}$$

With this first order formulation, one also needs to evolve the space derivatives. The simplest way of doing so is just to take the time derivative of Eq. (3) and interchange the order of space and time derivatives:

$$\begin{aligned}\partial_t A_k + \partial_r[-\beta^r A_k + \alpha\, Q\,\delta^r_k]\\= (2B_k^{\ r} - \alpha\, tr\, s\,\delta^r_k)A_r\,,\end{aligned} \tag{7}$$

$$\begin{aligned}\partial_t B_k^{\ i} + \partial_r[-\beta^r B_k^{\ i} + \alpha\, Q^i\,\delta^r_k]\\= (2B_k^{\ r} - \alpha\, tr\, s\,\delta^r_k)B_r^{\ i}\,,\end{aligned} \tag{8}$$

$$\begin{aligned}\partial_t D_{kij} + \partial_r[-\beta^r D_{kij} + \alpha\,\delta^r_k\,(K_{ij} - s_{ij})]\\= (2B_k^{\ r} - \alpha\, tr\, s\,\delta^r_k)D_{rij}\,,\end{aligned} \tag{9}$$

where we have noted

$$(\partial_t - \beta^r \partial_r) ln\,\alpha = -\alpha\, Q\,, \tag{10}$$
$$(\partial_t - \beta^r \partial_r)\beta^i = -2\alpha^2\, Q^i\,, \tag{11}$$
$$s_{ij} = (B_{ij} + B_{ji})/\alpha\,, \tag{12}$$

so choices of $Q$ and $Q^i$ will determine the gauge conditions.

The quantity $V_k$ introduced by Eq. (6) is very interesting. One can compute its time derivative from (9) but we will use the momentum constraint to transform that equation into

$$\begin{aligned}\partial_t V_k + \partial_r[-\beta^r\, V_k + \alpha\,(s^r_k - tr\, s\,\delta^r_k)]\\= \alpha\, G^0_{\ k} + \alpha A_r\,(K^r_{\ k} - tr\, K\,\delta^r_k)\\+ (D_{k\ r}^{\ s} - \delta^s_k\, D^{\ j}_{rj})\, K^r_{\ s} + (2B_k^{\ r} - \alpha\, tr\, s\,\delta^r_k)\, V_r\\- 2(D_{rk}^{\ \ s} - \delta^s_k\, D^j_{\ jr})(\alpha K^r_{\ s} - B^r_{\ s})\,.\end{aligned} \tag{13}$$

We will consider $V_k$ as an independent quantity to be evolved with Eq. (13). In that way, Eq. (6) becomes an algebraic constraint between $V_k$ and the spatial metric derivatives $D_{kij}$ (the algebraic form of the momentum constraint). This is crucial to ensure the hyperbolicity of the evolution system.

The set of Eqs. (1), (4), (7), (8), (9), and (13), together with the gauge Eqs. (10) and (11), has the special form we seek. The entire nonlinear system of the Einstein evolution equations can be written in the form of a first order system of balance laws as

$$\partial_t u + \partial_i F^i(u) = S(u) \tag{14}$$

which is familiar from many branches of physics. It is essential to point out that first spatial derivatives of the fields occur *only* through the flux term, and the source terms do not contain any derivatives. Note that we have *not* yet specified the time derivatives (10), (11) of the lapse nor the shift. Previously this form required the use of the harmonic slicing condition. This is one of the key results of this paper.

*Characteristic Fields.* We want to study under which conditions the evolution system is strictly hyperbolic. This point is crucial to apply advanced CFD numerical methods in which (the principal part of) the system is to be diagonalized by writing it in terms of the eigenfields which propagate along characteristic surfaces with their own characteristic speed.

When dealing with a first order system, one must first choose a fixed space direction to discuss hyperbolicity by considering only space derivatives along the selected direction. We will take the $k$ coordinate axis so, in the following, $k$ is never a dummy index.

¿From Eqs. (1), (10), and (11), it is clear that the coordinate normal lines will be characteristic (with speed $-\beta^k$). The corresponding characteristic fields are the lapse, the shift and metric components, and

$$A_{k'} , \quad B_{k'}^{\ i} , \quad D_{k'ij} \quad (k' \neq k) . \tag{15}$$

Also, the light cones are characteristic surfaces (speeds $-\beta^k \pm \alpha \sqrt{\gamma^{kk}}$). The corresponding eigenfields are

$$\gamma^{kk} (K_{ik'} - s_{ik'}) + \delta_i^k s_{k'}^{\ k} \mp \sqrt{\gamma^{kk}} \lambda_{ik'}^{\ k} \quad (k' \neq k) . \tag{16}$$

It is easy to see that both $tr K$ and $D_{k\ r}^{\ r}$ are algebraically independent from the fields listed above and can not be then recovered from them. These quantities, together with the vector $V_k$ and the set $Q$, $Q^i$, $A_k$, $B_k^{\ i}$ cannot be properly computed until one specifies a gauge condition to evolve the right-hand-side terms $Q$, $Q^i$ in (10,11). In that sense, the set of characteristic fields is not complete.

*A new class of gauge conditions.* We are interested in invariant slicing conditions. This means that the space-time slicing provided by our coordinate condition must be invariant under any transformation of the space coordinates of every slice. We must use then slicing scalars, like $\alpha$, $Q$ or $tr K$ and their proper time derivatives (note that the shift "vector" does not behave as a slicing vector; it is a vector under time independent transformations only). We also want to use an algebraic condition, because of their simplicity and low computational cost. If we restrict ourselves to zero order scalars, we can play only with $\alpha$ and we get either a geodesic slicing or one of its generalizations. If we allow also first order scalars, we get both $Q$ and $tr K$. The most general homogeneous algebraic condition is then

$$Q - f(\alpha) \, tr K = 0 , \tag{17}$$

where $f$ is an arbitrary function. The geodesic slicing is then included as a subcase with $f = 0$. The maximal slicing condition ($tr K = 0$) is included also as a limiting case when $f$ diverges. The $f = 1$ case corresponds to the harmonic slicing. Another interesting case is the "1+log" slicing [16,17], obtained when $f = 1/\alpha$; it mimics maximal slicing near a singularity, when the lapse collapses to zero. It has been very effective in evolving black hole spacetimes in 1D [16] and 3D [17]. The term "1+log" arises from the expression of $\alpha$ in terms of $\sqrt{\gamma}$ that one obtains when integrating Eq. (17) in the eulerian case (zero shift vector). Note however that the invariance of Eq. (17) ensures that one can apply it to obtain the same slicing even with a nonzero shift vector.

With these choices we get characteristic cones associated to the eigenfields

$$f \, tr K \, \gamma^{kk} + 2(s^{kk} - tr s \, \gamma^{kk}) \mp \sqrt{f \gamma^{kk}} \lambda_r^{kr} \quad (f \neq 0) , \tag{18}$$

with characteristic speeds

$$-\beta^k \pm \alpha \sqrt{f \, \gamma^{kk}} , \tag{19}$$

and we will call these *gauge speeds* because of their explicit dependence on the slicing condition. Gauge speed coincides with light speed only in the harmonic case ($f = 1$) and this may be considered to be the distinctive quality of harmonic slicing. Gauge speed becomes infinite for a maximal slicing, as one would expect with the maximal slicing elliptic equation. We note that in the maximal slicing case both $Q$ and $A_k$ are determined through the maximal slicing elliptic equation, and decouple from the other quantities. They are computed separately because Eqs. (17) and (18) provide now only one independent eigenvector ($tr K$). The discussion in this section applies then to the remaining set of equations which, as we show below, can be fully diagonalized.

The vector $V_k$ and the set $Q^i$, $B_k^{\ i}$, $D_{k\ r}^{\ r}$ still can not be obtained until one specifies a shift vector condition. The simplest choice is a zero shift vector ($Q^i = 0$, $B_{ij} = 0$) so that

$$V_i , \quad A_k - f \, D_{k\ r}^{\ r} \tag{20}$$

are the remaining eigenfields (zero speed). But one also has other choices of shift that still permit the system to be fully diagonalized [18]. The use of such shifts in numerical studies will be reported elsewhere.

It is clear that negative values of $f$ will lead to imaginary gauge speeds. Moreover, the set of eigenfields is complete only if $f \neq 0$. This means that the evolution system will be strictly hyperbolic iff $f > 0$. Note also that gauges with $f < 1$ will have poor singularity avoiding behavior because gauge speed would be lower than light speed. Therefore, cases with $f \geq 1$ will look more appealing for most Numerical Relativity applications, and many such choices have already been shown to work well for numerically evolved black hole spacetimes [16,17].

As a first test of these ideas we have applied this new formulation of the Einstein equations to numerical studies of spherical black hole spacetimes and compared results to those obtained with the standard ADM formulation of the equations [19] for a number of slicing conditions, including harmonic, "1+log", and maximal. In all cases the new numerical techniques allow the black hole to be evolved farther by orders of magnitude and with *much* higher accuracy. As a dramatic example of the power of this approach, in Fig. 1 we show the error in the evolution of the apparent horizon mass for the standard approach with staggered leapfrog compared with results from our code that uses the new formulation and advanced TVD methods [3]. Both results use identical resolution (200 grid zones) and maximal slicing. The ADM code crashes at $t = 150M$ after errors of more than 100% develop. The new approach can be run indefinitely with errors of a few percent. This formulation with advanced numerical techniques will be particularly important in the 3D case with black holes [17], apparent horizon boundary conditions [14,10,15], and gravitational waves [12]. Based on the encouraging results in spherical symmetry, a 3D code using these techniques is well under development, and results will be reported elsewhere.

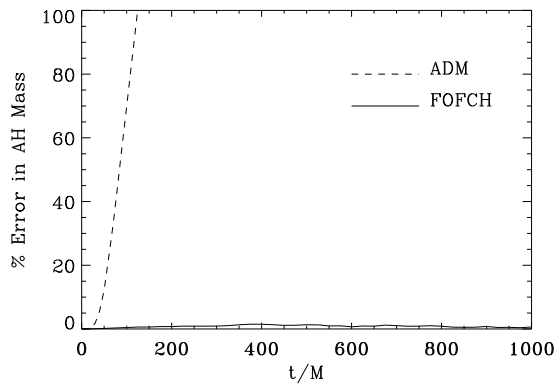

FIG. 1. Comparison between the standard ADM approach and the new FOFCH formulation. Evolution using the ADM approach cannot be continued beyond $t = 150M$, while the FOFCH continues indefinitely with a fraction of the error.

*Conclusion.* We have presented a powerful new first order, flux conservative, hyperbolic formulation of the Einstein equations that can be used with a wide class of gauge conditions, including maximal slicing. There are many advantages in this new formulation. First and foremost, it allows the use of advanced numerical methods that dramatically improve the accuracy and stability of numerical studies of the Einstein equations. Also, given that the system will be diagonalized in terms of eigenfields propagating along characteristic surfaces, one can specifically design boundary conditions for each of the fields, depending on the sign and value of their characteristic speeds (in fact, this is the idea behind many of the numerical methods). This is particularly important for the propagation of gravitational waves, as the techniques automatically identify and evolve the eigenfields that are propagating ("radiative variables"), as opposed to metric functions. This should also be crucial in the development of apparent horizon boundary conditions, as proper understanding of the inner boundary causal structure is required. The new class of algebraic slicings also gives insight into the singularity avoiding properties of general slicings and will allow strongly gravitating spacetimes. The analysis of the gauge speeds allows the use of proper causality conditions, and the study of potential problems when introducing new families of slicing gauges and shifts. The study of the structure of the remaining source terms will be important when the sources drive the evolution. Finally, this formulation allows the gravitational field to be treated on the same footing, with the same methods, as the equations of relativistic hydrodynamics. This combination should lead to a powerful approach to full GR hydrodynamics. As this formalism is already developed in the full 3D case, there is no difficulty in applying it there.

*Acknowledgements.* We would like to thank Larry Smarr for helpful discussions. This work is supported by the DGICyT of Spain under project PB91-0335. J.M. acknowledges a PFPI Fellowship from MEC of Spain. We also acknowledge the support of NCSA and NSF grants PHY94-07882, PHY/ASC93-18152 (Arpa supplemented), and INT94-14185.